\documentclass[twocolumn]{aastex631}
\usepackage{amsmath}
\usepackage{xcolor}

\newcommand{\lya}{Ly$\alpha$}
\newcommand{\lyb}{Ly$\beta$}
\newcommand{\xhi}{$x_{\rm HI}$}
\newcommand{\muv}{$M_{\rm UV}$}
\newcommand{\fesc}{$f_{\rm esc}$}
\newcommand{\ewlya}{$\textrm{EW}_{\textrm{Ly}\alpha}$}

\newcommand{\hi}{H\thinspace{\sc i}}
\newcommand{\hii}{H\thinspace{\sc ii}}

\received{March 8, 2023}
\revised{May 5, 2023}
\accepted{May 15, 2023}

\submitjournal{ApJL}

\begin{document}

\title{The Universe is at Most 88\% Neutral at $z=10.6$}

\author[0000-0001-8792-3091]{Sean Bruton}
\affiliation{Minnesota Institute for Astrophysics, University of Minnesota,\\
116 Church St SE, Minneapolis, MN 55455, USA}

\author[0000-0001-8792-3091]{Yu-Heng Lin}
\affiliation{Minnesota Institute for Astrophysics, University of Minnesota,\\
116 Church St SE, Minneapolis, MN 55455, USA}
\affiliation{School of Physics and Astronomy, University of Minnesota,\\ 116 Church St SE, Minneapolis, Minnesota 55455, USA}

\author[0000-0002-9136-8876]{Claudia Scarlata}
\affiliation{Minnesota Institute for Astrophysics, University of Minnesota,\\
116 Church St SE, Minneapolis, MN 55455, USA}
\affiliation{School of Physics and Astronomy, University of Minnesota,\\ 116 Church St SE, Minneapolis, Minnesota 55455, USA}

\author[0000-0001-8587-218X]{Matthew J. Hayes}
\affiliation{Stockholm University, Department of Astronomy and Oskar Klein Centre for Cosmoparticle Physics, AlbaNova University Centre, SE-10691, Stockholm,
Sweden}

\correspondingauthor{Sean Bruton}
\email{bruto012@umn.edu}



\begin{abstract}

Recent observations of GN-z11 with JWST have revealed a Ly$\alpha$ emission line with an equivalent width of 18$\pm 2$\AA. At $z=10.6$, this galaxy is expected to lie in the heart of reionization. We use a series of inhomogeneous reionization simulations to derive the distribution of the Ly$\alpha$ EW after traveling through the neutral intergalactic medium with varying average neutral gas fraction, $x_{HI}$. We use these distributions to place an upper limit of $x_{HI}$ $<$ 0.88 at $z=10.6$ at 95\% confidence level. We compare our upper limit to different reionization history models, which include the recently identified enhancement at the bright end of the luminosity function at $z>8$. We find that models in which faint galaxies have higher escape fraction compared to bright galaxies are favored by the new data.

\end{abstract}



\section{Introduction} \label{sec:intro}
In less than a year, $JWST$ has revolutionized our knowledge of the very-high-redshift universe, spectroscopically confirming and photometrically identifying galaxy candidates as early as only a few hundred million years after the Big Bang \citep{schaerer2022, brinchmann2022, finkelstein2022, curtis-lake2023, cameron2023, tang2023}.  Recently, \cite{bunker2023} presented the spectroscopic confirmation of the Lyman-break  galaxy candidate GN-z11. Using multiple rest-frame optical and UV emission lines, the redshift of GN-z11 was measured to be $z=10.603\pm 0.0013$, currently the highest known from emission lines for a likely star-forming galaxy (see discussion in \citet{bunker2023}). 

At redshift well above $10$, GN-z11 is expected to be embedded in a largely neutral universe, as inferred from a number of available data at $z\lesssim 8$ \citep[e.g.,][]{fan2006, mcgreer2011, mcgreer2015, ono2012, schroeder2013, choudhury2015, greig2016, mason2018, mason2019, greig2019, hoag2019, jung2020}. Inferring from lower-redshift observational studies suggests that the universe volume weighted neutral fraction (\xhi) is well above 90\% at $z\gtrsim 10$, although no observational constraints currently exist at such early epochs. The spectrum of GN-z11 offers the unique opportunity to understand the state of the diffuse hydrogen at $z>10$. Indeed, the spectrum reveals the presence of a strong \lya\ emission line, with a rest-frame equivalent width (EW) of $18\pm 2$\AA, and a velocity shift of the \lya\ with respect to systemic velocity of $\Delta v_{Ly\alpha}\approx 550$km~s$^{-1}$. This implies that GN-z11 resides in a an ionized bubble, perhaps not surprising given that it is one of the brightest galaxies known at $z>10$.

\lya\ emission has been used extensively to place constraints on reionization. \lya\ is a resonant transition in neutral hydrogen and is expected to completely disappear from the spectra of high-redshift galaxies when these are embedded in a fully neutral intergalactic medium \citep[IGM; ][]{dijkstra2014, dijkstra2017}. The evolution of the absolute number density of \lya\ emitters has been used to constrain the timeline of reionization  \citep[e.g.,][]{rhoads2001, malhotra2004, malhotra2006, hu2019, wold2021, morales2021a, ning2022}. However, because \lya\ is the byproduct of star-formation, its disappearance cannot be uniquely  interpreted as signaling a completely neutral IGM. Studies of the reionization history of the universe are now often based on relative properties, e.g., on the relative fraction of \lya\ emitters among star-forming galaxies \citep{stark2010, pentericci2011, jung2018}, on the \lya\ equivalent width distribution \citep[\ewlya; ][]{mason2018a, hoag2019, whitler2020}, or on the clustering of \lya\ emitters \citep{ouchi2010, sobacchi2015, ouchi2018, yoshioka2022}. Additionally, the environment in which a galaxy resides has an impact on the visibility of \lya. A galaxy located within a region of the universe previously ionized by either itself, or a previous generation of stars, i.e., residing in what are commonly referred to as ``ionized bubbles,''  would be more likely to show \lya\ in emission because  \lya\ photons would have time to redshift out of resonance before encountering the neutral IGM. The probability of observing \lya\ emission then  depends on the size of the ionized bubbles and the \lya\ velocity shifts with respect to the IGM.

In this paper, we use the observations of GN-z11 to constrain \xhi\ at $z=10.6$. Specifically, we use a set of simulations to forward model the evolution of the \lya\ EW distribution in the presence of an increasingly neutral IGM and use these distributions to construct the likelihood of observing a galaxy with the observed \ewlya\ and \muv\ properties as GN-z11. 

The structure of the paper is as follows. In Section~\ref{sec:model}, we postprocess 21cmFast simulations to create a model of Lyman-$\alpha$ emitters (LAEs). We then present our inference model. In Section~\ref{subsec:results}, we explore the implication of our result for the reionization history and discuss how the result depends on the assumptions. Finally we conclude in Section~\ref{sec:conclusion}. We assume a $\Lambda$CDM cosmology with $H_0=$ 67.66 km s$^{-1}$ Mpc$^{-1}$ \citep{planck-collaboration2020}.

\section{Modeling} \label{sec:model}
The end goal of our modeling is to make an inference on the global neutral fraction of the universe, \xhi, at $z=10.6$, in light of the fact that we observe one LAE with \ewlya$=18\pm 2$\AA. To estimate \xhi, we will use a hierarchical Bayesian approach, similar to the one used in \citet{bruton2023} and \citet{mason2018a}. We will populate a series of large inhomogenous reionization simulations with LAEs and calculate the  distribution of the \ewlya\ after transmission through the IGM of different ionization fractions. 

\subsection{Populating a Reionization Simulation with LAEs}
Our model is built on the backbone of 21cmFastv2 \citep{mesinger2007, mesinger2011, mesinger2016}, a seminumerical code that combines excursion set formalism and perturbation theory to create simulations of the universe throughout reionization. We use nine simulations with a volume average neutral fraction varying between  \xhi=0.01 to \xhi=0.92, each $4.1\rm{Gpc}^3$ in volume. When modeling the evolution of \xhi, 21cmFast takes into account recombinations, photoheating star-formation suppression, supernova feedback, and radiation. For each dark matter halo of mass, $M_h$, 21cmFast computes the integrated \lya\ optical depth as a function of the velocity offset from line center. The simulations naturally account for the evolving presence of ionized bubbles around galaxies of different masses and residing in different environments.  

In what follows, we define the emergent \lya\ ($\rm Ly\alpha_{emer}$), as the \lya\ line luminosity escaping from the interstellar and circumgalactic medium of a galaxy, while the \lya\ observed after passing through the IGM is referred to as transmitted \lya\ ($\rm Ly\alpha_{tran}$).
To assign an emergent  \lya\ to a dark matter halo, we follow \citet{bruton2023}. Briefly, after using the $M_h$-\muv\ relation from \citet{mason2015} to assign an \muv\ magnitude to each halo, we use the \ewlya\ - \muv\ probability distribution from \citet{mason2018}, calibrated with \citet{debarros2017} observations, to randomly draw the emergent \ewlya for each halo. This is discussed in more detail in the following paragraph. The emergent \lya\ line profile shape is assumed to be a truncated Gaussian, with velocity offset and Full-Width-Half-Maximum (FWHM) dependent on \lya\ luminosity. The velocity offset scales linearly with the log of the \lya\ luminosity (see \citet{bruton2023} for more details), and we set FWHM equal to velocity offset, which is consistent with the findings in   \citet{verhamme2018}, wherein they fit the relation between observed low- and high-z LAEs' velocity offsets and FWHMs. They find the empirical relation to be consistent with the one-to-one relation predicted from radiation transfer modeling. We also note that \citet{hayes2023a} find, through a Bayesian hierarchical inference model, that GN-z11's intrinsic red wing \lya\ emission has a velocity offset of 400 km s$^{-1}$  and a FWHM of 433 km s$^{-1}$ , consistent with the one-to-one. Finally, we use the line profiles, together with the velocity dependent optical depths of \lya, to calculate the \lya\ transmission,  $T= \frac{\rm Ly\alpha_{tran}}{\rm Ly\alpha_{emer}}$, and the 
transmitted \ewlya\ as $EW^{\rm  tran}_{\rm Ly\alpha} = T\times EW^{\rm emer}_{\rm  Ly\alpha}$. 

\begin{figure}[t]
\centering
\includegraphics[width=.99\linewidth]{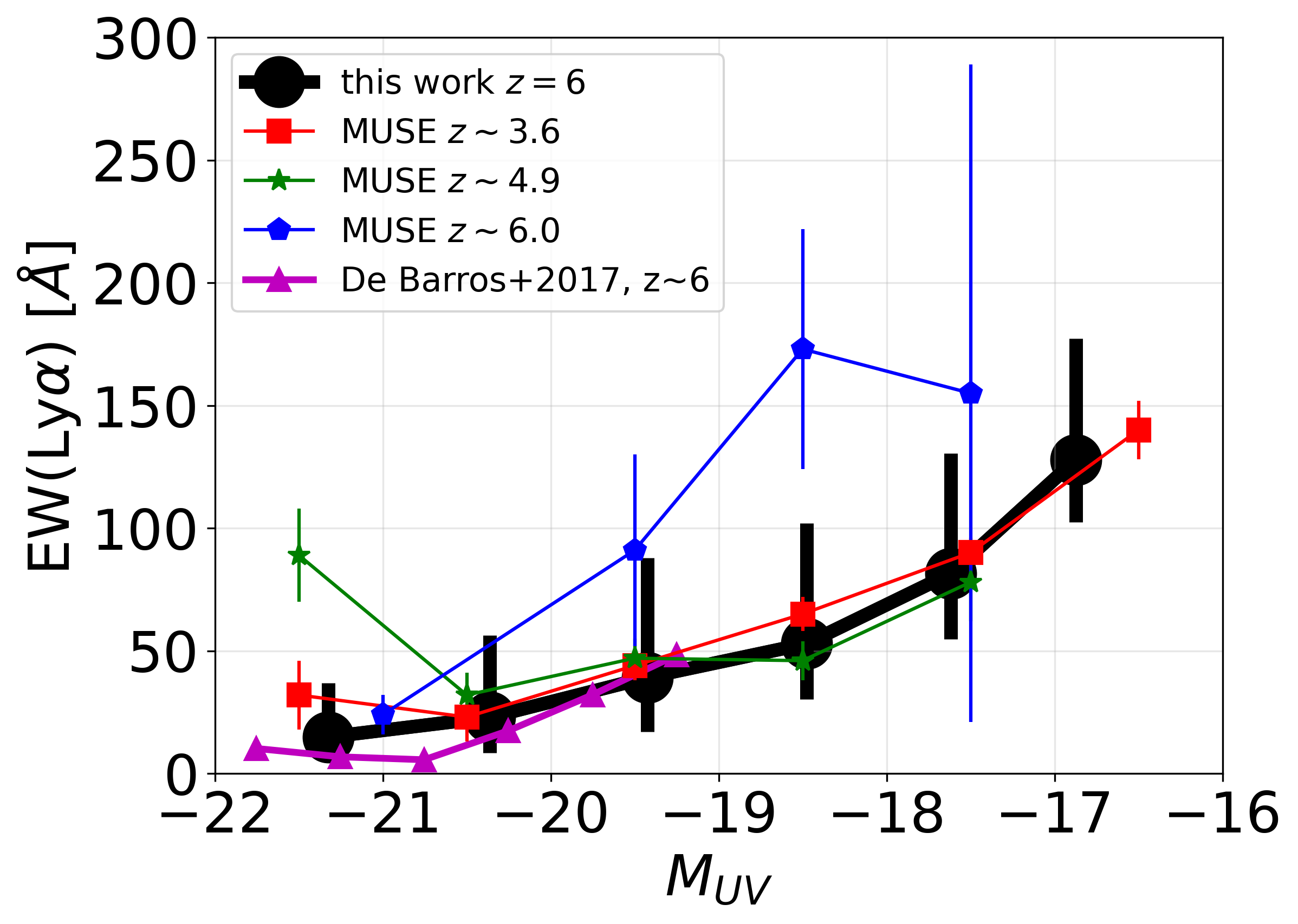}
\caption{EW(\lya) as a function of \muv. The red squares, green stars, and blue pentagons are the measurements from the MUSE HUDF survey at redshift $z=$3.6, 4.9, and 6.0, respectively. The magenta triangles are the median measurements from \citet{debarros2017}. The black circles are the results from our model, after matching the MUSE HUDF survey flux limit.
 \label{fig:ew_dist} }
\end{figure}

Our analysis rests on the assumption that the distribution of emergent \ewlya\ at \muv$= -21.5$ does not change with redshift.  
For a generic \muv, this distribution takes the form, from \citet{mason2018}, 

\begin{equation}
\begin{aligned}
p(\textrm{EW}^{\rm emer}_{\textrm{Ly}\alpha}\mid M_{\textrm{UV}}) = {} & \frac{A(M_{\textrm{UV}})}{W_c(M_{\textrm{UV}})}e^{-\frac{\textrm{EW}_{\textrm{Ly}\alpha}}{W_c(M_{\textrm{UV}})}}H(\textrm{EW}_{\textrm{Ly}\alpha})\\ & + [1-A(M_{\textrm{UV}})]\delta(\textrm{EW}_{\textrm{Ly}\alpha})
\label{ew_distribution}
\end{aligned}
\end{equation}
where $H(\textrm{EW}_{\textrm{Ly}\alpha})$ is the Heaviside step function and $\delta({\textrm{EW}_{\textrm{Ly}\alpha}})$ is the Dirac delta function. The parameter $A$ accounts for the fraction of galaxies that do not emit \lya\ and $W_c$ determines the exponential decline of the probability distribution function toward larger \ewlya. Both parameters are functions of \muv; the general behavior is that brighter galaxies are more likely to be nonemitters and have a stronger exponential cutoff so that UV bright galaxies are less likely to have large \ewlya. In Figure~\ref{fig:ew_dist} we show the MUSE Hubble Ultra Deep Field (HUDF) Survey measurements of the average emergent \ewlya\ at different redshifts as a function of \muv\ \citep{hashimoto2017}, $z\sim6$ observations from \citet{debarros2017}, and the prediction of our model. We apply an observed flux lower limit $>8\times10^{-19}$erg~s$^{-1}$ for \lya\ to match the selection limit in the MUSE survey (the flux limit in the \citet{debarros2017} data is similar at $2.2\times10^{-18}$erg~s$^{-1}$ across the entire wavelength range, deeper in wavelength ranges without skylines).

We limit to data to $z\lesssim 6$, to limit the impact of the transmission through the IGM and isolate the effects of the interstellar and circumgalactic medium. This permits a comparison between \ewlya\ distributions at different redshifts to see if emergent \ewlya\ is independent of redshift. The comparison in Figure \ref{fig:ew_dist} confirms that there is no evidence of evolution in the median emergent \ewlya\ from $z=3.6$ to $z=4.9$, and weak evidence of evolution from $z=4.9$ to $z=6.0$, a redshift range corresponding to 800 million years. We note that a partially neutral IGM at $z=6$ would remove LAEs with weak \ewlya\ from the observations and thus bias the distribution to be higher, and that the universe may not be fully ionized by $z=6$ \citep{qin2021}. 

Still, we test the impact of the intrinsic \ewlya\ distribution by adopting a different exponential cutoff strength, $W_c$, for our intrinsic \ewlya\ distribution and redo our inference. For \muv$=-21.5$, $W_c=19$, as fit from the \citet{debarros2017} data. As a test, we weaken the exponential cutoff, setting $W_c=43$, which, allowing galaxies to take on \ewlya\ values greater than the GN-z11 value of 18\AA\ more easily, decreases the probability of having a galaxy with an \ewlya\ of 18\AA\ by $\approx$20\%. When redoing the inference with this new distribution, we infer the same 95\% upper limit on \xhi. This is an extremely different intrinsic \ewlya\ distribution, but the impact on the inferred value of \xhi\ is negligible. This shows that even if there is strong evolution in the intrinsic \ewlya\ distribution toward $z=10.6$ allowing galaxies to have larger \ewlya, its impact on our result will be small. 

\begin{figure}[t]
\centering
\includegraphics[width=.99\linewidth]{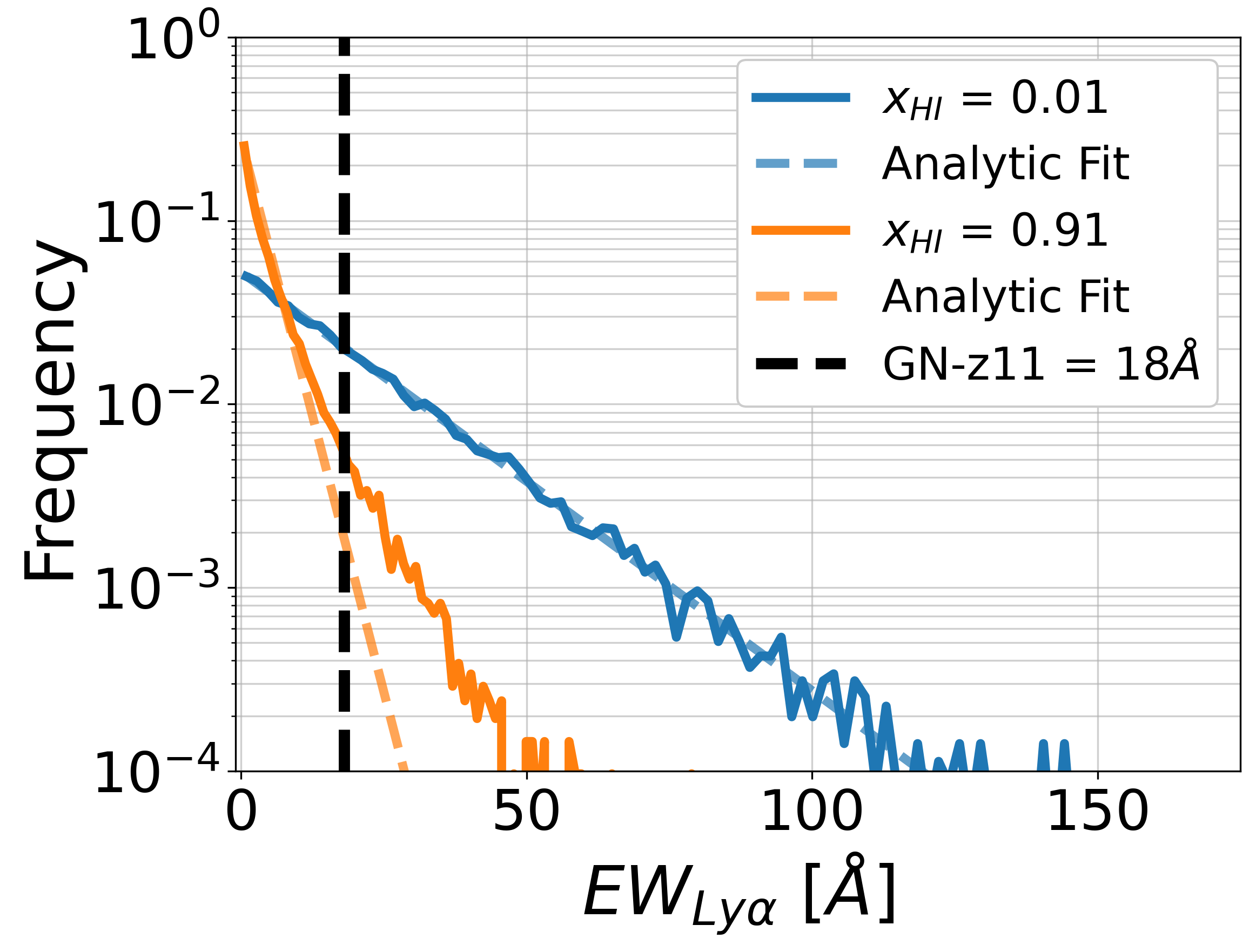}
\caption{The probability of EW(\lya) for varying values of \xhi, given that the galaxy has \lya\ emission (i.e. excluding nonemitters). The solid lines are empirically taken from the simulation after IGM attenuation--the dashed lines are an attempt to fit the first term of Eq. \ref{ew_distribution} to the distribution. The fit fails when the universe is partially neutral because there is excess probability of having high \ewlya\ compared to an exponential cutoff.
 \label{fig:ew_prob} }
\end{figure}

The \lya\ transmission through an inhomogenous IGM, particularly in the case of a highly neutral IGM, changes the \ewlya\ distribution, which is no longer well described by Equation \ref{ew_distribution}. This is demonstrated in Figure \ref{fig:ew_prob}, where we show $p(\textrm{EW}^{\rm tran}_{\textrm{Ly}\alpha}\mid M_{\textrm{UV}}=-21.5,x_{HI})$ for two values of \xhi. For the highly neutral universe with \xhi$=0.9$, the probability distribution function (PDF) of the transmitted \ewlya\ can no longer be modeled with the first term in Eq.~ \ref{ew_distribution}. The observed excess of high-EW LAEs compared to the prediction from the ionized universe is a direct result of the inhomogeneity of the reionization process. Some galaxies reside in ionized bubbles, and their $\rm{Ly}\alpha_{emer}$ is not very attenuated by the IGM. This results in a distribution of \ewlya\ that does not follow an exponential decline. This also highlights an important aspect of our simulations--the LAEs reside in a variety of local neutral fractions, which may differ from the global neutral fraction. However, when these LAEs are taken in conglomerate, the probability of having a given \ewlya\ is dependent on the global neutral fraction, not the local neutral fraction. Thus, the inference arising from this probability yields the global value, rather than a local value. Seeing as we have only one object to feed into our inference, we expect the posterior to be weakly constrained; more objects, probing more lines of sight, would be much more constraining.

\subsection{Hierarchical Bayesian Inference of $x_{\rm{HI}}$}

Using Bayes' theorem, we write the \xhi\ posterior 

\begin{multline*}
p(x_{\rm{HI}} \mid EW^{\rm obs}_{Ly\alpha},M^{-21.5}_{UV}) \propto \\
\int p(EW^{\rm obs}_{Ly\alpha} \mid EW^{tran}_{Ly\alpha}, x_{\rm{HI}}, M^{-21.5}_{UV})\\ 
p(x_{\rm{HI}}) p(EW^{tran}_{Ly\alpha} \mid M^{-21.5}_{UV}) {\rm d} EW^{tran}_{Ly\alpha}
\end{multline*}
where $p(EW^{obs}_{Ly\alpha} \mid EW^{tran}_{Ly\alpha},x_{\rm{HI}}, M^{-21.5}_{UV})$ is the likelihood of the measured \ewlya\ given the neutral fraction, and marginalizing over the values of $EW^{tran}_{\rm Ly\alpha}$.  $p(x_{\rm{HI}})$ is the prior on the global neutral fraction, which we assume to be uniform between [0,1].  $p(EW^{tran}_{Ly\alpha} \mid M^{-21.5}_{UV})$  is the PDF of the transmitted EW computed in the previous section for \muv$=-21.5$. 
\lya\ optical depths vary smoothly with \xhi, so we interpolate between our nine simulations to build a function for the \ewlya\ distributions dependent on \xhi\ and \muv. We assume that the likelihood of $EW^{obs}_{Ly\alpha}$ is a normal, with known standard deviation that we set equal to the measurement uncertainty provided by \cite{bunker2023}.

We use a Metropolis-Hastings algorithm to sample the posterior, accounting for measurement errors on \ewlya. 

\section{Results and Discussion} \label{subsec:results}

 \begin{figure*}[t]
\centering
\includegraphics[width=.45\linewidth]{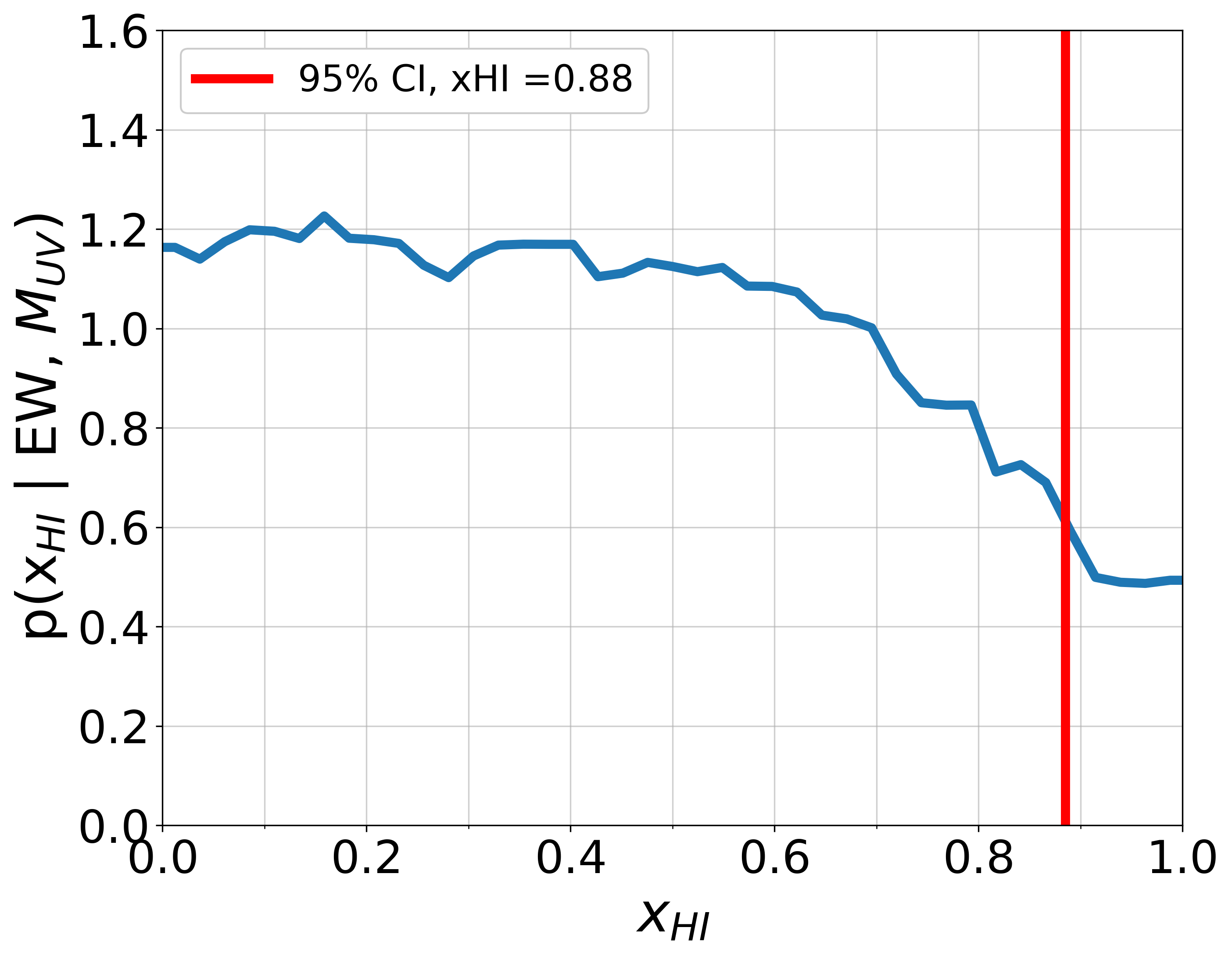}
\includegraphics[width=.45\linewidth]{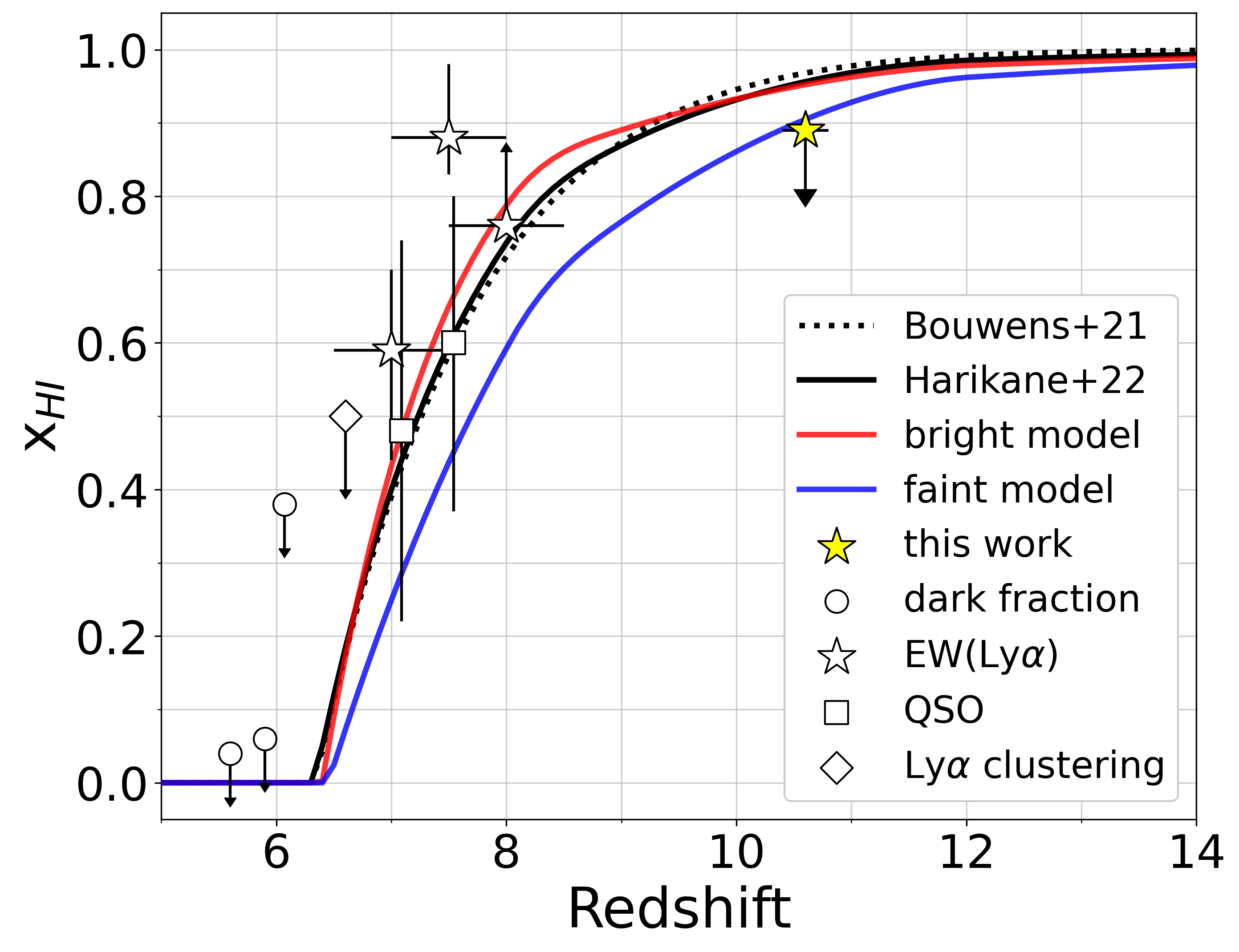}
\caption{ $Left$: the posterior distribution of \xhi.  $Right$: the reionization history. The dash and solid black lines are constant \fesc\ models with Schechter LF \citep{bouwens2021} and double power-law LF \citep{harikane2023}. The red and blue lines are the models where bright and faint galaxies have higher escape fraction, respectively. The observational constraints are shown in white markers, and the constraint in constraint inferred by GN-z11 is marked as yellow star.  
 \label{fig:history} }
\end{figure*}

The main result of our analysis is presented in Figure~\ref{fig:history}, where we show the  \xhi\  posterior in the left panel. Not surprisingly, with only one galaxy, the value of \xhi\ is poorly constrained. However, it is clear that \xhi\ is smaller than 0.88 at 95\% confidence level (the 95\% credibility region is indicated by the vertical line in Figure~\ref{fig:history}). In the right panel of Figure~\ref{fig:history}, we show the constraint on \xhi\ derived from GN-z11 on the reionization history timeline, and compare it with  constraints on the neutral fraction from other observations: 
\lya\ EW of galaxies \citep{mason2018,mason2019, hoag2019}; 
the clustering of \lya\ emitter galaxies \citep{ouchi2010,greig2016}; 
\lya\ and \lyb\ dark fraction \citep{mcgreer2015}; 
QSO damping wings \citep{davies2018}. While the GN-z11 observation does not constrain the absolute value on \xhi\ (one can just as easily say that \xhi\ is $>0.04$ with 95\%), the fact that the observation is very far back in the expected timeline of reionization offers a lever, allowing one to distinguish between reionization models.

Predicting the redshift evolution of the volume averaged neutral fraction depends on the number of ionizing sources present at each time, their ionizing spectrum, and the ability of ionizing radiation to escape into the IGM. Once these are known, $x_{\text{\hi}}$ can be computed by solving:

\begin{equation} \label{eq:reionization}
\frac{d (1 - x_{\text{\hi} })}{dt} = \frac{\dot{N}_{ion}}{n_{\text{H}}} - \frac{( 1- x_{\text{\hi}})}{t_{\text{rec}}}, 
\end{equation}
\citep{madau1999, robertson2013, ishigaki2018} where $\dot{N}_{ion}$ is the ionizing photon production rate, $n_{\text{H}}$ is the comoving gas number density, and $t_{\text{rec}}$ is the recombination time scale.  $n_{\text{H}}$ and $t_{\text{rec}}$ are defined as 
\begin{equation}
n_{\text{H}} = \frac{ X_p\Omega_b \rho_{c} }{m_{\text{H}}}, 
\end{equation}
\begin{equation}
t_{\text{rec}} = [ C_{\text{\hii}}~\alpha_B(T) (1 + Y_p/4X_p) n_{\text{H}} (1+z)^3  ]^{-1}, 
\end{equation}
where $X_p$, $Y_p$ are the primordial mass fraction of hydrogen and helium, $\Omega_b$ is the baryon energy density fraction, $\rho_{c}$ is the critical density, ($\Omega_b \rho_{c} = 4.2\times 10^{-31}$ [$g/cm^3$]), $C_{\text{\hii}} \equiv$ $\langle n_{\text{\hii}^2} \rangle$/$\langle n_{\text{\hii}}\rangle^2$ is the clumping factor, and $\alpha_B(T)$ is the case B recombination coefficient. Here we assume $C_{\text{\hii}} = 3$, and $\alpha_B=2.6\times10^{-13}$ cm$^3$s$^{-1}$, for an electron temperature of $10^4$K. We solve Equation~\ref{eq:reionization} iteratively, assuming the boundary condition that $x_{\text{\hi}}=1.0$ at $z=18$, i.e., we assume that the first sources of ionizing photons appear at this redshift.

The major uncertainties in the calculation of \xhi\ above are introduced in the ionizing photon production rate, $\dot{N}_{ion}$, that can be expressed as the product of three components: 
\begin{equation}
\dot{N} =f_{esc}\,\xi_{ion}\,\rho_{UV},
\end{equation}
where \fesc\ is the absolute LyC escape fraction, $\xi_{ion}$ is the ionizing photon production efficiency, and $\rho_{UV}$ is the UV luminosity density, i.e., the integral of the galaxy UV luminosity function (LF).

The UV LF has changed very quickly after JWST observations. Early $JWST$ results indicate that bright galaxies at $z>8$ are more numerous than previously expected  \citep{finkelstein2022}, and a double power-law LF (as opposed of a classical Schechter LF) was proposed to account for  the newly discovered population of bright galaxies \citep{bouwens2021,harikane2023}.  
Are these new bright sources responsible for the observed neutral fraction at $z=10.6$? The relative contribution of bright and faint galaxies to the reionization budget is unconstrained. Which objects prevail has an impact on the reionization timeline: reionization  starts late but completes rapidly if  bright galaxies dominate, \citep{sharma2016, naidu2020}, while it starts early but proceeds slowly if faint galaxies played the major role \citep{finkelstein2019}.

In what follows we consider four simple reionization history models that include the recent developments in the LF studies and  vary the contribution of bright and faint galaxies to the ionizing budget.  
For all models, we assume log($\xi_{ion}$/[Hz erg$^{-1}$])=25.7, characteristic of high-redshift star-forming galaxies \citep{ning2023,tang2023}. 

First, we consider the reionization models that assume different luminosity functions at $z>8$: 
a Schechter LF \citep[dashed line,][]{bouwens2021}, and a double power-law LF \citep[solid line;][]{harikane2023}. At $z\leq8$, both models are calculated with the Schechter LF in \citet{bouwens2021}.  We adopt a constant \fesc$=$0.12 at all redshifts in both models.  
We find that although there are more UV bright ($M_{UV}<-22$) galaxies when the double power-law  LF is used, the early ($z>9$) reionization histories are roughly equivalent in the two models. Neither model, however, is able to reionize the universe early enough to account for the \xhi\  upper limit at $z=10.6$, as in both models \xhi\ reaches 90\% only by $z_{90} \approx 9.4$.

Next, we consider the models where bright (\muv\ $ < -18$) or faint galaxies (\muv\ $\geq -18$) dominate the ionizing photon budget.  To achieve this goal we vary the escape fraction as a function of both  \muv\ and redshift: 

\[
    \text{\fesc} = 
    \begin{cases}
    0.08 + 0.022(z-6) -0.12(M_{UV} + 18),\:\: \textrm{bright}\\
    0.08 + 0.022(z-6) + 0.02(M_{UV} + 18),\:\: \textrm{faint}
    \end{cases}
\]

\noindent
with flattening at \fesc$=$0.5. We refer to these models as bright and faint models. In both models, we adopt the DP LF from \citet{harikane2023} at $z>8$ and the Schechter LF from \citet{bouwens2021} at $z\leq8$. 

The bright model (red curve in Figure~\ref{fig:history}) shows a late and rapid reionization similar to the models with constant \fesc, despite the fact that the dominant contributors to the ionizing photon budget are different (in the constant $f_{esc}$ models, faint galaxies dominate because of their high space density compared to bright galaxies). In the faint model (blue curve in Figure ~\ref{fig:history}), the ionizing photon contribution of galaxies with \muv$>-18$ is enhanced with respect to the constant \fesc\ models, causing reionization to start earlier. Specifically, we find that  $x_{\text{\hi}}$ reaches 90\%  by $z_{90} \approx 10.9$ and progresses more slowly compared to models where brighter galaxies dominate. 
The faint model is the preferred scenario for the early stage of reionization ($z>10$) in light of the \xhi$<0.88$ constraint inferred from GN-z11.

An independent constraint on the reionization history comes from observations of the cosmic microwave background. Recently, {\it Planck} \citep{planck-collaboration2020} measured the (integrated) optical depth to Thomson scattering to be $\tau = 0.056\pm0.007$, ruling out the need of a large contribution from  galaxies at $z>10$, and favoring a fast and late reionization process. The redshift evolution of the free electron fraction, $x_e(z)$, constrained by {\it Planck}, however, depends on the specific form used to model the reionization history, as $\tau$ is only sensitive to the integral of the electron density over time, and not its detailed shape. The FlexKnot model assumed in PlanckCollaboration~VI limits the contribution to the optical depth of $z>15$ galaxies to less than 1\%, but leaves space to the possibility that the universe was not fully neutral at $z=10.6$, with a free electron fraction of $\approx 8-10$\%, consistent with the upper limit presented here.

\section{Conclusions}\label{sec:conclusion}
At a record distance of $z=10.6$, GN-z11 is the highest-redshift \lya-emitting galaxy known, well within the heart of the reionization epoch. We use an inhomogeneous reionization simulation to derive the probability distribution of the transmitted \ewlya\ through the IGM as a function of \muv\ and the average neutral gas fraction, \xhi. We use these distributions to estimate the posterior distribution function on \xhi\ at $z=10.6$. With data for only one galaxy, we place an upper limit on the global neutral fraction at $z=10.6$, i.e., \xhi$\lesssim 0.88$ with a probability of 95\%.  With this constraint we are able to exclude reionization histories dominated by bright galaxies; a scenario wherein faint galaxies have a higher escape fraction of ionizing photons, and so drive reionization, is favored.

%

\vspace{5mm}


\textit{Software:} \texttt{Astropy} \citep{robitaille2013}, \texttt{SciPy} \citep{oliphant2007}, \texttt{NumPy} \citep{vanderwalt2011}, and \texttt{Matplotlib} \citep{hunter2007}.



\bibliography{HIfraction}{}
\bibliographystyle{aasjournal}



\end{document}